\documentstyle[prl,aps,epsf,floats]{revtex}
\begin{document}
\draft
\wideabs{
\title{MAGNETIC-FIELD-INDUCED HYBRIDIZATION OF ELECTRON SUBBANDS IN A
COUPLED DOUBLE QUANTUM WELL}
\author{V.T.~Dolgopolov, G.E.~Tsydynzhapov, A.A.~Shashkin,
E.V.~Deviatov}
\address{Institute of Solid State Physics, Chernogolovka, Moscow
District 142432, Russia}
\author{F.~Hastreiter, M.~Hartung, A.~Wixforth}
\address{Ludwig-Maximilians-Universit\"at, Geschwister-Scholl-Platz
1, D-80539 M\"unchen, Germany}
\author{K.L.~Campman, A.C.~Gossard}
\address{Materials Department and Center for Quantized Electronic
Structures, University of California, Santa Barbara, California
93106, USA}
\maketitle
\begin{abstract}
We employ a magnetocapacitance technique to study the spectrum of the
soft two-subband (or double-layer) electron system in a parabolic
quantum well with a narrow tunnel barrier in the centre. In this
system unbalanced by gate depletion, at temperatures $T\agt 30$ mK we
observe two sets of quantum oscillations: one originates from the
upper electron subband in the closer-to-the-gate part of the well and
the other indicates the existence of common gaps in the spectrum at
integer fillings. For the lowest filling factors $\nu=1$ and $\nu=2$,
both the common gap presence down to the point of one- to two-subband
transition and their non-trivial magnetic field dependences point to
magnetic-field-induced hybridization of electron subbands.
\end{abstract}
}

\thispagestyle{myheadings}
\markboth{}{\hfil to be published in JETP Letters}
A soft two-subband electron system, or double electron layer, is the
simplest system having a degree of freedom, which is associated with
the third dimension, in the integer (IQHE) and fractional (FQHE)
quantum Hall effect. As compared to a conventional two-subband
electron system with vanishing distance $d$ between electron density
maxima, such as the one in single heterojunctions, in the double
layer the energy spacing between subbands is very sensitive to
intersubband electron transfer because of large $d\agt
a_B=\varepsilon\hbar^2/me^2$. Since pioneering papers
\cite{boeb,suen} much attention is paid to the investigation of
balanced systems with symmetric electron density distributions.
While in this case the origin of the IQHE at even integer filling
factors is trivial, the symmetric-antisymmetric level splitting
caused by tunneling gives rise to the IQHE at odd integer filling
factors. The absence of certain IQHE states at low odd integer
fillings \cite{boeb,suen} was interpreted in Refs.~\cite{mac,brey} as
due to the Coulomb-interaction-induced destruction of
symmetric-antisymmetric splitting in strong magnetic fields.
Observation was reported of the bilayer many-body IQHE at filling
factor $\nu=1$ \cite{murphy,lay} and FQHE
\cite{suen1,eis,suen2,suen3} whose origin, alternatively, was
attributed to interlayer correlation effects
\cite{chak,yosh,fertig,he}. The case of an unbalanced system with
strongly asymmetric electron density distributions was studied in
Ref.~\cite{davies}. At relatively high filling factors the authors
\cite{davies} observed an interplay between "single- and double-layer
behaviour" and explained this in terms of charge transfer between two
electron subbands without appealing exchange and correlation effects.

Here, using a capacitive spectroscopy method, we investigate the
spectrum of two-dimensional electrons at a quantizing magnetic field
in a parabolic quantum well that contains a narrow tunnel barrier for
the electron systems on either side. In the gate-depletion-unbalanced
double-layer system, new gaps with unusual magnetic field dependences
have been detected at filling factors $\nu=1$ and $\nu=2$. We argue
that these emerge as a result of magnetic-field-induced hybridization
of electron subbands.

The sample is grown by molecular beam epitaxy on semi-insulating GaAs
substrate. The active layers form a 760~\AA\ wide parabolic well. In
the center of the well a 3 monolayer thick Al$_x$Ga$_{1-x}$As
($x=0.3$) sheet is grown which serves as a tunnel barrier between
both parts on either side. The symmetrically doped well is capped by
600~\AA\ AlGaAs and 40~\AA\ GaAs layers. The sample has two ohmic
contacts (each of them is connected to both electron systems in two
parts of the well) and a gate on the crystal surface with area
$120\times 120$ $\mu$m$^2$. The presence of the gate electrode
enables us both to tune the carrier density in the well and to
measure the capacitance between the gate and the well. For
capacitance measurements we apply an ac voltage $V_{ac}=2.4$~mV at
frequencies $f$ in the range 3 to 600~Hz between the well and the
gate and measure both current components as a function of gate bias
$V_g$, using a home-made $I-V$ converter and a standard lock-in
technique. Our measurements are performed in the temperature interval
between 30~mK and 1.2~K at magnetic fields of up to 16~T.

\begin{figure}[t]
\epsfxsize=\columnwidth
\epsffile{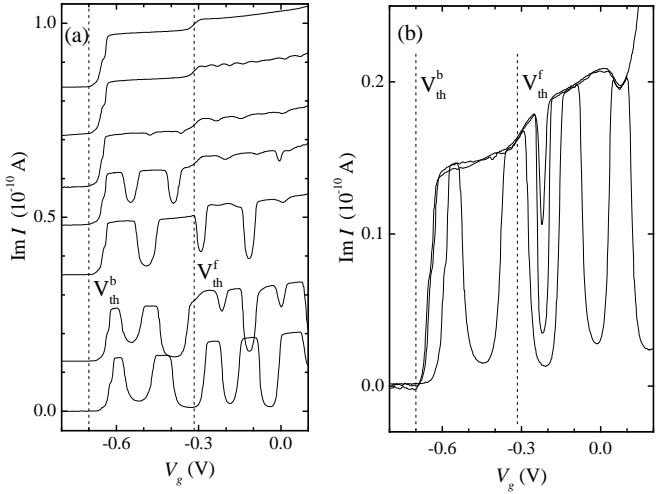}
\caption{Dependence of the imaginary current component on gate
voltage at $f=110$~Hz at different temperatures and magnetic fields:
(a) $B=0,0.67,1.34,1.84,2.51,3.68,4.34$~T at $T=30$~mK, the lines are
shifted proportionally to steps in $B$; (b) $T=30,620,880$~mK at
$B=6$~T.\label{exp}}
\end{figure}

The dependences of the imaginary current component on gate voltage at
different magnetic fields are shown in Fig.~\ref{exp}(a). In zero
magnetic field at $V_{th}^b=-0.7\mbox{ V} <V_g< V_{th}^f=-0.31\mbox{
V}$ electrons fill only one subband in the back part of the well,
relative to the gate. With increasing $V_g> V_{th}^f$ a second
electron subband starts to collect electrons in the front part of the
well, which is indicated by an increase of the capacitance. In
magnetic fields of about 1.3~T at low temperatures we observe two
sets of quantum oscillations: first, the oscillations at $V_g>
V_{th}^f$ are due to the modulation of the thermodynamic density of
states in the upper electron subband. They are typical of a three
electrode system (see, e.g., Refs.~\cite{ash,dens}) and depend only
weakly on temperature in the regime investigated. Second, the
oscillations at $V_g< V_{th}^f$ originate from the conductivity
oscillations in the lower electron subband and so these are
accompanied by peaks in the real current component. With increasing
the magnetic field one more set of oscillations emerges formed by
additional minima at $V_g> V_{th}^f$ (Fig.~\ref{exp}(a)). The small
values of capacitance at the oscillation minima as well as the
non-zero active current component reflect that the conductivity
$\sigma_{xx}$ vanishes for both electron subbands. Since related to
$\sigma_{xx}$, these common oscillations are strongly
temperature-dependent, whereas the measured capacitance in between
the deep minima depends weakly on temperature. As seen from
Fig.~\ref{exp}(b), the weak oscillations reflecting the thermodynamic
density of states in the upper subband persist after the appearance
of the common oscillations. In particular, these do not change at
all, when located between deep minima, as the latter develop. For
coincident positions, the minima for the two kinds of oscillations
trigger with changing temperature.

\begin{figure}[t]
\centerline{\epsfysize=6.98cm
\epsffile{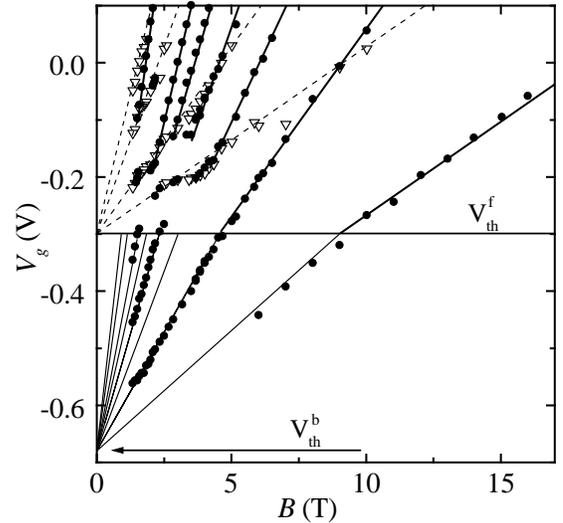}}
\caption{Landau level fan chart as found from the minima of the
density of states in the upper electron subband (open symbols, dashed
lines) and of the conductivity of the lower electron subband (filled
symbols, solid lines) and of the conductivity of the double-layer
electron system (filled symbols, bold lines). The bold line
disruptions signify the absence of common gaps. At crossing points of
the dashed and bold lines the corresponding minima trigger with
changing temperature.\label{fan}}
\end{figure}

Fig.~\ref{fan} presents a Landau level fan diagram in the ($B,V_g$)
plane for our sample. Positions of the density of states minima in
the upper electron subband are shown by open symbols. These minima
correspond to the filling factors $\nu_2=1,2,4,6$ in the upper
subband. The conductivity minima are marked in the figure by solid
symbols. In the gate voltage interval $V_{th}^b <V_g< V_{th}^f$ we
see the filling factors $\nu_1=1,2,4,6$ in the lower subband.
Furthermore, for $V_g> V_{th}^f$ the common oscillations define the
third Landau level fan. The straight lines of this fan are parallel
to those for the upper electron subband and correspond to
$\nu=1,2,3,4,5,6,8,10$ which is the filling factor as determined by
the electron density $N_s$ in the quantum well. In the ($B,V_g$)
plane the different fan line slopes below and above $V_g=V_{th}^f$
(Fig.~\ref{fan}) correspond to the capacitance values before and
after the jump near $V_g=V_{th}^f$ (Fig.~\ref{exp}). One can see from
Fig.~\ref{fan} that despite with varying the gate voltage $V_g>
V_{th}^f$ the electron density changes essentially in the front part
of the well as indicated by the fan line slopes, it is the integer
$\nu$ at which common gaps are observed in the double-layer system.

\begin{figure}[t]
\centerline{\epsfysize=7cm
\epsffile{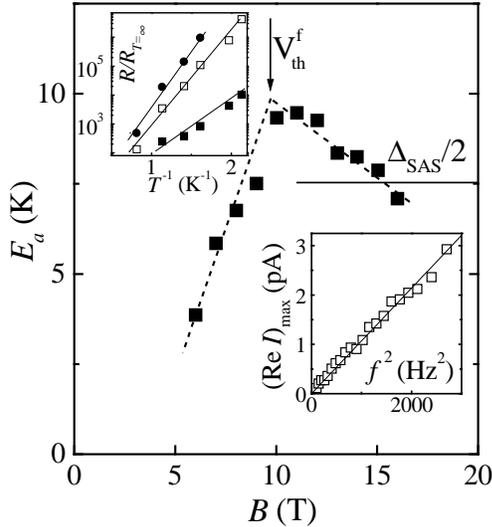}}
\caption{Activation energy as a function of magnetic field at filling
factor $\nu=1$. The insets display the frequency dependence of the
active current component peak (bottom) and Arrhenius plot of the peak
amplitude for $B=6,10,14$~T (top).\label{activ}}
\end{figure}

The activation energy in the common oscillation minima is found from
the temperature dependence of peaks in the active current component,
which accompany capacitance minima. In the limit of vanishing active
current component the peak amplitude is expected to be proportional
to $f^2\sigma_{xx}^{-1}$. To make sure that the measuring frequency
is sufficiently low, we investigate the frequency dependence of the
active current component, see the bottom inset to Fig.~\ref{activ}.
In the frequency range where the above relation holds, the activation
energy is simply determined from Arrhenius plot of the peak amplitude
(the top inset to Fig.~\ref{activ}). Fig.~\ref{activ} displays the
magnetic field dependence of the activation energy for filling factor
$\nu=1$. This dependence is quite non-trivial: the activation energy
is a maximum at about $V_g=V_{th}^f$, where a second electron subband
starts to be filled, and then it monotonically decreases with
magnetic field up to the balance point. A similar behaviour is found
also for filling factor $\nu=2$. Although the gaps at filling factors
$\nu >2$ are also maxima near the threshold voltage $V_{th}^f$, at
higher fields, unlike the gaps at $\nu=1$ and $\nu=2$, they vanish in
some intervals of $B$ (or $V_g$). This is indicated by disruptions of
the fan lines in Fig.~\ref{fan}.

\begin{figure}[t]
\centerline{\epsfxsize=6.22cm
\epsffile{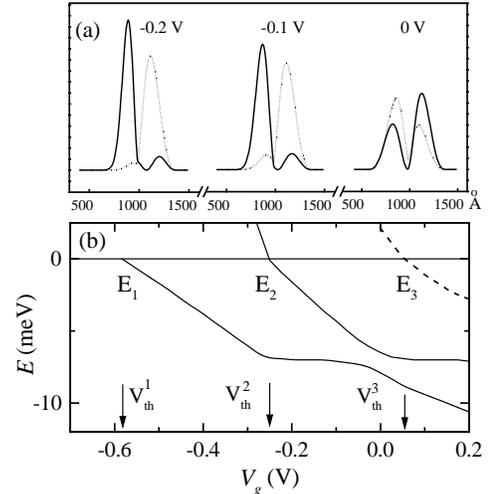}}
\caption{Calculated at $B=0$, electron density distributions for the
two lower energy bands in the quantum well (a) and positions of the
energy band bottoms as a function of gate voltage (b). The dotted and
solid lines in case (a) correspond to $E_1$ and $E_2$,
respectively.\label{band}}
\end{figure}

The band structure of our sample in the absence of magnetic field is
known from far-infrared spectroscopy and magnetotransport
investigations on samples fabricated from the same wafer
\cite{hart,ens}. It agrees with the result of self-consistent Hartree
calculation of energy levels in a coupled double quantum well
(Fig.~\ref{band}). In the calculation one reasonably assumes that all
electron subbands have a common electrochemical potential which
corresponds to the zero-point on the energy scale in
Fig.~\ref{band}(b). In agreement with experiment, only one energy
band is occupied by electrons in the range $V_{th}^1 <V_g< V_{th}^2$,
two energy bands are filled at $V_{th}^2 <V_g< V_{th}^3$ and three of
them are filled above $V_g=V_{th}^3$. The band splitting at a zero
gate voltage is symmetric-antisymmetric splitting
$\Delta_{SAS}=1.3$~meV. Fig.~\ref{band}(a) shows the electron density
profiles for the two lower energy bands in the quantum well at three
different gate voltages. We note that for both energy bands, even far
from the balance, the wave function is not completely localized in
either part of the quantum well.

Experimentally, the possibility of all electrons collecting in one
part of the quantum well (so-called broken-symmetry states
\cite{jung}) is excluded because of the co-existence of the Landau
level fan for the upper subband and the one determined by common
oscillations (Fig.~\ref{fan}).

One can tentatively expect that the experimental data find their
interpretation in terms of relative shift of Landau level ladders
corresponding to two electron subbands. At fixed integer $\nu$, the
conductivity $\sigma_{xx}$ of a bilayer system should tend to zero in
the close vicinities of Landau-level-fan crossing points in the
($B,V_g$) plane, at which both individual filling factors
$\nu_f,\nu_b$ are integers, as long as the Fermi level remains in a
gap between quantum levels for two electron subbands. Obviously, in
between the crossing points the common gap closes as soon as the
Fermi level pins to both of the quantum levels. Such a behaviour is
indeed observed in the experiment at filling factors $\nu >2$, see
Fig.~\ref{fan}. We note that the presence (absence) of common gaps
was identified in Ref.~\cite{davies} as "single (double) layer
behaviour". In contrast, for a conventional two-subband electron
system with vanishing distance between electron density maxima,
common gaps at integer $\nu$ are expected to close in negligibly
narrow intervals on the Landau-level-fan lines where both quantum
levels from two electron subbands cross the Fermi level.

However, such simple considerations fail to account for the common
gaps at filling factors $\nu=1,2$ which do not disappear in the
entire range from the threshold $V_{th}^f$ to the balance point
(Figs.~\ref{fan},\ref{activ}). We explain this behaviour as a result
of magnetic-field-induced hybridization of the wave functions of two
electron subbands, which gives rise to the creation of new gaps in
the bilayer spectrum.

In a soft two-subband electron system, quantum level energies for two
Landau level ladders can get equal only if the corresponding wave
functions are orthogonal, i.e., if the Landau level numbers are
different. Apparently, this is not the case for $\nu=1,2$ as well as
for higher $\nu\neq 4m$ ($m$ is integer) near the balance point. The
absence of orthogonality implies that the bilayer system is described
by the hybrid wave function that is a linear combination of the wave
functions of two electron subbands. The appearance of new gaps, as a
result, is crucially determined by intersubband charge transfer in
magnetic field to make the band bottoms coincident. We note that this
process is impossible in the conventional two-subband system as
discussed above. Although in our soft two-subband system the distance
between electron density maxima (Fig.~\ref{band}) is close to the
in-plane distance between electrons, the charge transferred is
estimated to be small. This is confirmed experimentally by the
absence of appreciable deviations of the data points from the
upper-subband-fan lines (Fig.~\ref{fan}) as determined by
zero-magnetic-field capacitance at $V_g> V_{th}^f$ (Fig.~\ref{exp}).
It is clear that the magnetic-field-induced hybridization generalizes
the case of symmetric electron density distributions corresponding to
formation of $\Delta_{SAS}$. From the first sight it seems natural to
expect that the common gaps at $\nu=1,2$ decrease with magnetic field
and approach $\Delta_{SAS}$ at the balance point (Fig.~\ref{activ}).
Yet, for all filling factors in question the situation is far more
sophisticated because the spin splitting, which is comparable to the
hybrid splitting, comes into play. The bilayer spectrum, then, is
determined by their competition which, in principle, may even lead to
closing common gaps in some intervals of magnetic field. For example,
at $\nu=2$ the actual gap is given by the splitting difference and so
it zeroes for equal splittings. In our experiment, for the simplest
case of $\nu=1$ one can expect that over the range of magnetic fields
used the many-body enhanced spin gap is large compared to
$\Delta_{SAS}$ \cite{dens}. That stands to reason, it is the smaller
splitting that corresponds to $\nu=1$ (Fig.~\ref{activ}). For $\nu=2$
the very similar behaviour of the gap with magnetic field hints that
at these lower fields the hybrid splitting is dominant. As a result
of interchange of the hybrid and spin splittings, odd $\nu >1$ near
the balance in our case correspond to the spin rather than hybrid
gaps.

In summary, we have performed magnetocapacitance measurements on a
bilayer electron system in a parabolic quantum well with a narrow
tunnel barrier in its centre. For asymmetric electron density
distributions created by gate depletion in this soft two-subband
system we observe two sets of quantum oscillations. These originate
from the upper electron subband in the front part of the well and
from the gaps in the bilayer spectrum at integer fillings. For the
lowest filling factors $\nu=1$ and $\nu=2$, the common gap formation
is attributed to magnetic-field-induced hybridization of electron
subbands, dependent on the competition between the hybrid and spin
splitting.

We gratefully acknowledge J.P.~Kotthaus, A.V.~Chaplik and M.~Shayegan
for fruitful discussions of the results. This work was supported in
part by Volkswagen-Stiftung under Grant No.~I/71162, Deutsche
Forschungsgemeinschaft, AFOSR under Grant No.~F49620-94-1-0158, INTAS
under Grant No.~93-933, the Russian Foundation for Basic Research
under Grant No.~97-02-16829, and the Programme "Statistical Physics"
from the Russian Ministry of Sciences. The Munich - Santa Barbara
collaboration has been also supported by a joint NSF-European Grant
and the Max-Planck research award.

\end{document}